# Direct Imaging Discovery of a Jovian Exoplanet Within a Triple Star System

Kevin Wagner[1 *], Dániel Apai[1,2], Markus Kasper[3], Kaitlin Kratter[1], Melissa McClure[3], Massimo Robberto[4], Jean-Luc Beuzit[5,6]

[1]Department of Astronomy and Steward Observatory, The University of Arizona, 933 N. Cherry Avenue, Tucson, AZ 85721, USA.

[2]Lunar and Planetary Laboratory, The University of Arizona, 1640 E. University Boulevard, Tucson, AZ 85718, USA.

[3]European Southern Observatory, Karl-Schwarzschild-Strasse 2, D-85748 Garching, Germany.

[4]Space Telescope Science Institute, 3700 San Martin Drive, Baltimore, MD 21218, USA.

[5]Université Grenoble Alpes, Institut de Planétologie et d'Astrophysique de Grenoble, F-38000 Grenoble, France.

[6]Centre National de la Recherche Scientifique, Institut de Planétologie et d'Astrophysique de Grenoble, F-38000 Grenoble, France.

*Correspondence to: kwagner@as.arizona.edu

**Abstract**: Direct imaging allows for the detection and characterization of exoplanets via their thermal emission. We report the discovery via imaging of a young Jovian planet in a triple star system and characterize its atmospheric properties through near-infrared spectroscopy. The semi-major axis of the planet is closer relative to that of its hierarchical triple star system than for any known exoplanet within a stellar binary or triple, making HD 131399 dynamically unlike any other known system. The location of HD 131399Ab on a wide orbit in a triple system demonstrates that massive planets may be found on long and possibly unstable orbits in multi-star systems. HD 131399Ab is one of the lowest mass (4±1 $M_{Jup}$) and coldest (850±50 K) exoplanets to have been directly imaged.

**One Sentence Summary:** This article presents the discovery and characterization of a dynamically active young exoplanet within a triple star system.

**Main Text:**

Thousands of planets around other stars have been discovered (e.g. *1, 2*), revealing a greater diversity than predicted by traditional planet formation models based on the solar system. Extreme examples are planets within binary and multiple star systems, which form and evolve in variable radiation and gravitational fields. Direct imaging allows for the detection and characterization through spectroscopy of long-period giant planets – enabling constraints to be placed on planet formation models via predictions of planet population statistics and atmospheric properties (*3*). However, most direct imaging surveys have traditionally excluded visual binary or multiple systems whose separations are less than a few hundred astronomical units (au) due to the assumption that such planetary systems would either be disrupted or never form, as well as the increased technical complexity of detecting a planet amongst the scattered light of multiple stars. As a result of this observational bias, most directly imaged exoplanets have been found around single stars.

Since multi-star systems are as numerous as single stars (*4*), building a complete census of long-period giant planets requires investigation of both configurations. In principal, planets on wide orbits (detectable by direct imaging) might arise more frequently in multi-star systems due to planet-planet or planet-star interactions (*5, 6*). Such interaction could even produce planets on chaotic orbits that wander between the stars (*7, 8*). To investigate the frequency of long-period giant planets both around single stars and in multi-star systems, we are sampling a population of ~100 young single and multiple A-type stars in the nearby Upper Scorpius-Centaurus-Lupus association using the Very Large Telescope (VLT) and the Spectro-Polarimetric High-Contrast Exoplanet Research instrument (SPHERE; *9*). Here we report the discovery of the first planet detected in our on-going survey, and the widest-orbit planet within a multi-star system.

**Observations and Discovery of HD 131399Ab**

HD 131399 (HIP72940) is a triple system (*10*) in the 16±1 Myr old Upper Centaurus-Lupus association (UCL; *11-13*) at a distance of 98±7 pc (*14*) whose basic properties are given in Table 1. The system's membership in UCL is confirmed by its parallax and kinematics (*11-13*), and the well-constrained age of the association provides greater confidence in the young age of the system than for most directly imaged exoplanet host stars (see supplementary online text for the detailed age analysis). Despite its youth, the system shows no evidence of infrared excess and thus its primordial disk has likely been depleted to beneath detectable levels (*15*).

We observed HD 131399 on June 12, 2015, obtaining a wide range of near-infrared spectral coverage from *Y*– to *K*-band (0.95–2.25 μm) and diffraction-limited imaging with an 8.2-meter telescope aperture. Our observations (*10*) resulted in the discovery of HD 131399Ab, a point source with a contrast to HD 131399A of $10^{-5}$ and projected separation of 0.84 arcseconds, or 82±6 au (see Fig. 1 and Table 1). After the initial discovery, we obtained follow-up observations (*10*) to verify whether the faint source is physically associated with the parent star (i.e. shares common proper motion) and to improve the quality of the near-infrared spectrum, enabling the characterization of the planet's atmospheric properties.

| | **HD 131399A** | **HD 131399Ab** | **HD 131399B** | **HD 131399C** |
|---|---|---|---|---|
| Spectral Type | A1V[1] | T2–T4 | G | K |
| Mass | 1.82 $M_{\odot}$[2] | 4±1 $M_{Jup}$ | 0.96 $M_{\odot}$[2] | 0.6 $M_{\odot}$ |
| Temperature ($T_{eff}$) | 9300 K | 850±50 K | 5700 K | 4400 K |
| Projected separation from A (arcsec) | | 0.839±0.004 *2015 June*<br>0.834±0.004 *2016 March*<br>0.830±0.004 *2016 May* | 3.149±0.006 *2015 June*<br>3.150±0.006 *2016 March*<br>3.149±0.006 *2016 May* | 3.215±0.006 *2015 June*<br>3.220±0.006 *2016 March*<br>3.220±0.006 *2016 May* |
| Position angle (Degrees E of N from A) | | 194.2±0.3 *2015 June*<br>193.8±0.3 *2016 March*<br>193.5±0.3 *2016 May* | 221.9±0.3 *2015 June*<br>221.5±0.3 *2016 March*<br>221.8±0.3 *2016 May* | 222.0±0.3 *2015 June*<br>221.9±0.3 *2016 March*<br>222.1±0.3 *2016 May* |
| *J* magnitude | 6.772±0.018 | 20.0±0.2 | | |
| *H* magnitude | 6.708±0.034 | 19.7±0.2 | | |
| *K*-band magnitude | *K*=6.643±0.026 | *K1*=19.1±0.1 | *K1*=8.5±0.1 | *K1*=10.5±0.1 |

**Table 1. Basic parameters of the stars and directly imaged planet in HD 131399.** The mass, effective temperature, and spectral type of the previously unresolved B and C (except where noted) were estimated from their *K1* luminosity (*17-19, 35*). The planet's temperature and spectral type were determined through spectral fitting (see next section on characterization). Apparent *J*, *H*, and *K* magnitudes for HD 131399A were obtained from (*36*). References: 1 (*37*), 2 (*38*).

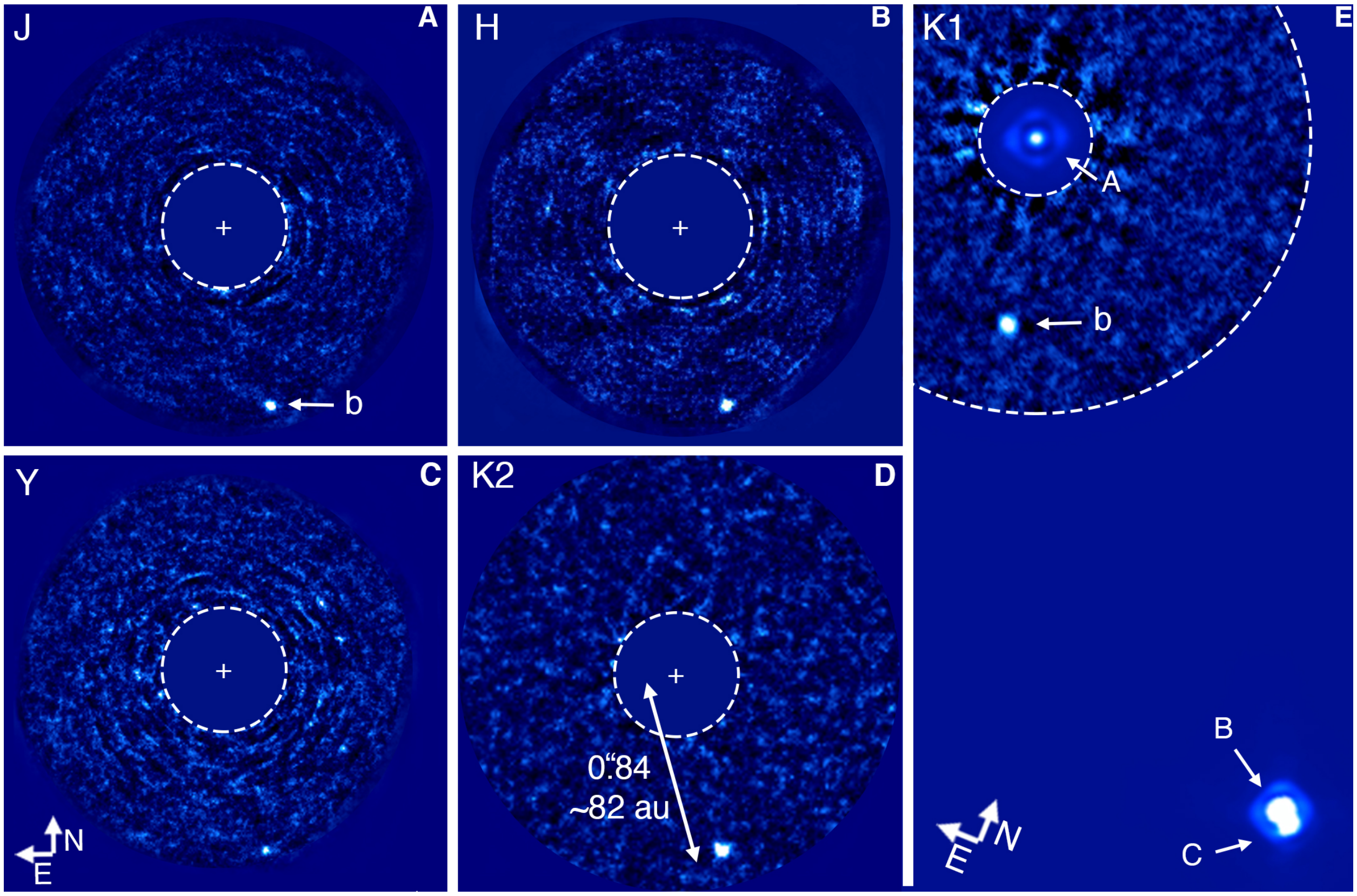

**Fig. 1. Near-infrared VLT/SPHERE images of HD 131399Ab and the hierarchical triple star system HD 131399ABC.** The central regions that are affected by the coronagraph and residual scattered starlight are blocked by a mask (dashed) in panels A-D, with the location of star A indicated by the crosshairs. Panel E shows a composite of the PSF-subtracted region (dashed) superposed on the wide field *K1* image showing the stellar components of the system, whose luminosities are adjusted to the level of the planet for clarity. In each image the luminosity of component A (but not B and C) has been suppressed by the use of a coronagraph. Panels A-C were processed with angular and spectral differential imaging to subtract the stellar PSF, while panel D and the PSF-subtracted region of panel E were processed only with angular

differential imaging (*10*). Panels A-D share the same field orientation.

We detect HD 131399Ab with a signal to noise ratio in *Y* (1.04 μm), *J* (1.25 μm), *H* (1.62 μm), *K1* (2.11 μm) and *K2* (2.25 μm) of 9.3, 13.2, 15.5, 23.5 and 11.9, respectively. Following astrometric calibrations (*10*), we measure a positional displacement to HD 131399A of $\Delta\alpha$ (right ascension) = 12±8 mas and $\Delta\delta$ (declination) = 6±8 milliarseconds (mas) over the eleven-month baseline, where the uncertainties are dominated by the calibration of the instrument orientation across the two epochs. This allows us to reject the hypothesis of a background object, which would have moved relative to HD 131399A by $\Delta\alpha$ = 27.3±0.6 mas and $\Delta\delta$ = 28.8±0.6 mas due to the relatively high proper motion of the system (*14*). Assuming a Keplerian orbit for the planet with a semi-major axis equivalent to its projected separation of 82 au yields a period of roughly 550 years, which for a face-on circular orbit over eleven months is expected to produce ~9 mas of relative motion, which is consistent with the observations.

The bound planet hypothesis is also supported by the low probability of detecting an unbound object within UCL that happens to share a similar spectral type to HD 131399Ab (as discussed in the next section). Following the arguments in (*16*), the false alarm rate of an unassociated objected with a planet-like spectrum per field of view is $\sim 2\times10^{-7}$. The total false alarm probability of one such object appearing in our 33 fields of view (so far explored in our survey) is given by the binomial distribution, resulting in a probability of $\sim 6.6\times10^{-6}$. Although the probability of detecting a bound giant planet is not yet well established, results from the first several hundred stars surveyed suggest this is around a few percent – orders of magnitude higher than the probability of detecting an unbound object with a planet-like spectrum.

**Characterization of HD 131399Ab**

We convert the planet's *J–, H–, and K1*-band aperture photometry to a mass estimate via comparison to widely used evolutionary tracks for "hot-start" initial conditions (*16-18*), in which the planet retains its initial entropy of formation. Systematic interpolation between hot-start evolutionary tracks yields a mass of 4±1 $M_{Jup}$, which places HD 131399Ab firmly in the planetary mass regime. Even in the unlikely event that the system is much older (a few hundred Myr) companion Ab would necessarily be planetary mass (<13 $M_{Jup}$). The H vs. H-K color of the planet is inconsistent with the cold-start scenario, in which the planet has lost some fraction of its initial entropy due to inefficient accretion (*20*), while consistent with hot-start models including a partly cloudy atmosphere and/or super-solar metallicity (Fig. S1).

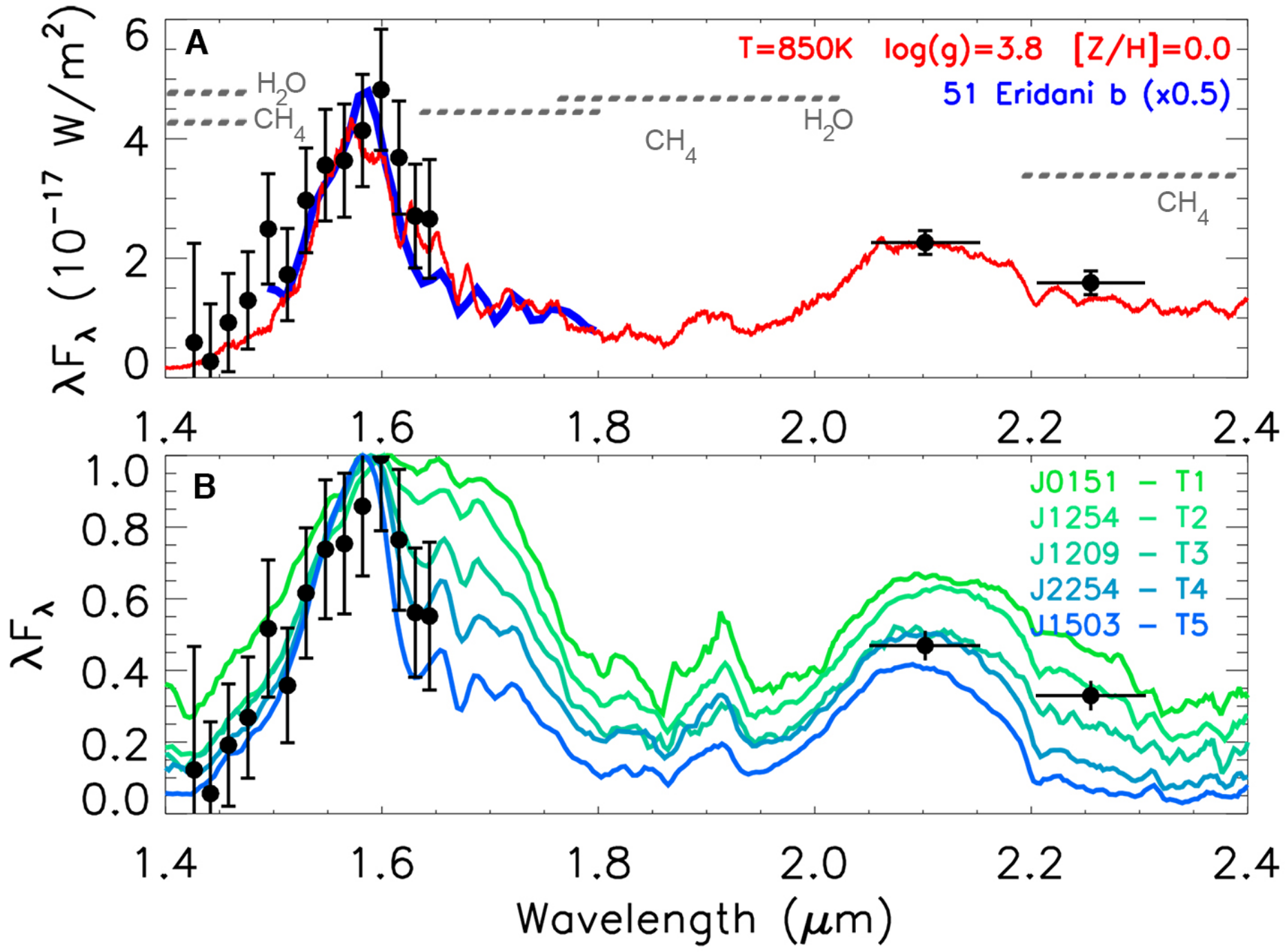

**Fig. 2: Near-infrared spectrum of HD 131399Ab.** Panel A: HD 131399Ab spectrum (black) alongside the best-fit model atmosphere in red (*18*), with $T_{eff}$ = 850 K and *log(g)* = 3.8 cm/s$^2$, showing water and methane absorption in the atmosphere with the approximate absorption

regions indicated by the dashed lines. The spectrum of the T-type exoplanet 51 Eri b (*16*) is shown in blue, scaled by 50% to roughly match the luminosity of HD 131399Ab. Panel B: Near-infrared spectrum of HD 131399Ab and spectra of standard field brown dwarfs (*39, 40*), with each 1.4–2.4 μm spectrum normalized independently in $\lambda F_\lambda$ units (equivalent to power per unit area). The objects' labels correspond to the 2MASS object designations (J2000 hours and minutes of right ascension) and the spectral type.

Using the integral field spectrograph (IFS, *21*) on SPHERE we obtained a 0.95–1.65 μm spectrum. This spectrum allows the characterization of water and methane absorption bands within 1.4–1.6 μm, while the signal to noise ratio is too poor in the individual spectral channels at shorter wavelengths to be useful in spectral analysis. In *K*-band, where the contrast with the star is more favorable, the dual-band images also probe the 2.2 μm methane absorption. Like the exoplanet 51 Eridani b (*16*) and other field (non-exoplanet) T-type brown dwarfs, the near-infrared spectrum of HD 131399Ab (Fig. 2) displays prominent methane and water absorption bands. The data are in good agreement with models of exoplanetary atmospheres (*18*), allowing us to estimate the atmospheric properties of effective temperature and surface gravity. Systematic exploration of interpolated atmospheric models indicates $T_{\rm eff}$ = 850±50 K and *log(g)* = $\mathbf{3.8^{+1.7}_{-0.8}}$ (cm/s$^2$), where the uncertainty in surface gravity is mostly dominated by systematic uncertainties within the models (namely in the cloud properties) and not by the model-data fit. Comparison to standard classifications of field brown dwarfs (Fig. 2, panel B) indicates a spectral type of T2–T4.

The transition between L and T spectral types ($T_{\rm eff}$ ~2100–1300 K and $T_{\rm eff}$ ~1300–600 K, respectively) is marked by the appearance of *H*– and *K*-band methane absorption in the atmospheres of the cooler T dwarfs. In *J* vs. *J–H* color-magnitude space (Fig. 3) this appears as bluer color (more negative *J–H* color) compared to the hotter L dwarfs. At the threshold of the L/T transition the photosphere becomes brighter in the *J*-band, as silicate clouds transition from above to below the photosphere (*22*). The fact that cloudy directly imaged exoplanets (such as HR8799bcde, β Pic b, or 2M1207b) appear at the bottom of the L-dwarf sequence argues for cloud layers in these low-gravity objects that are thicker than in their higher-gravity brown dwarf counterparts (*23, 24*). In contrast, HD 131399Ab and the two other directly imaged T-type exoplanets follow the T-dwarf sequence, which we interpret as evidence for a similarity between the mostly or fully cloud-free atmospheres of these exoplanets and cool field brown dwarfs. HD 131399Ab is the closest directly imaged exoplanet to the L/T transition, which is consistent with the partly cloudy atmosphere suggested by the *H* vs. *H-K* hot-start model predictions (Fig. S1).

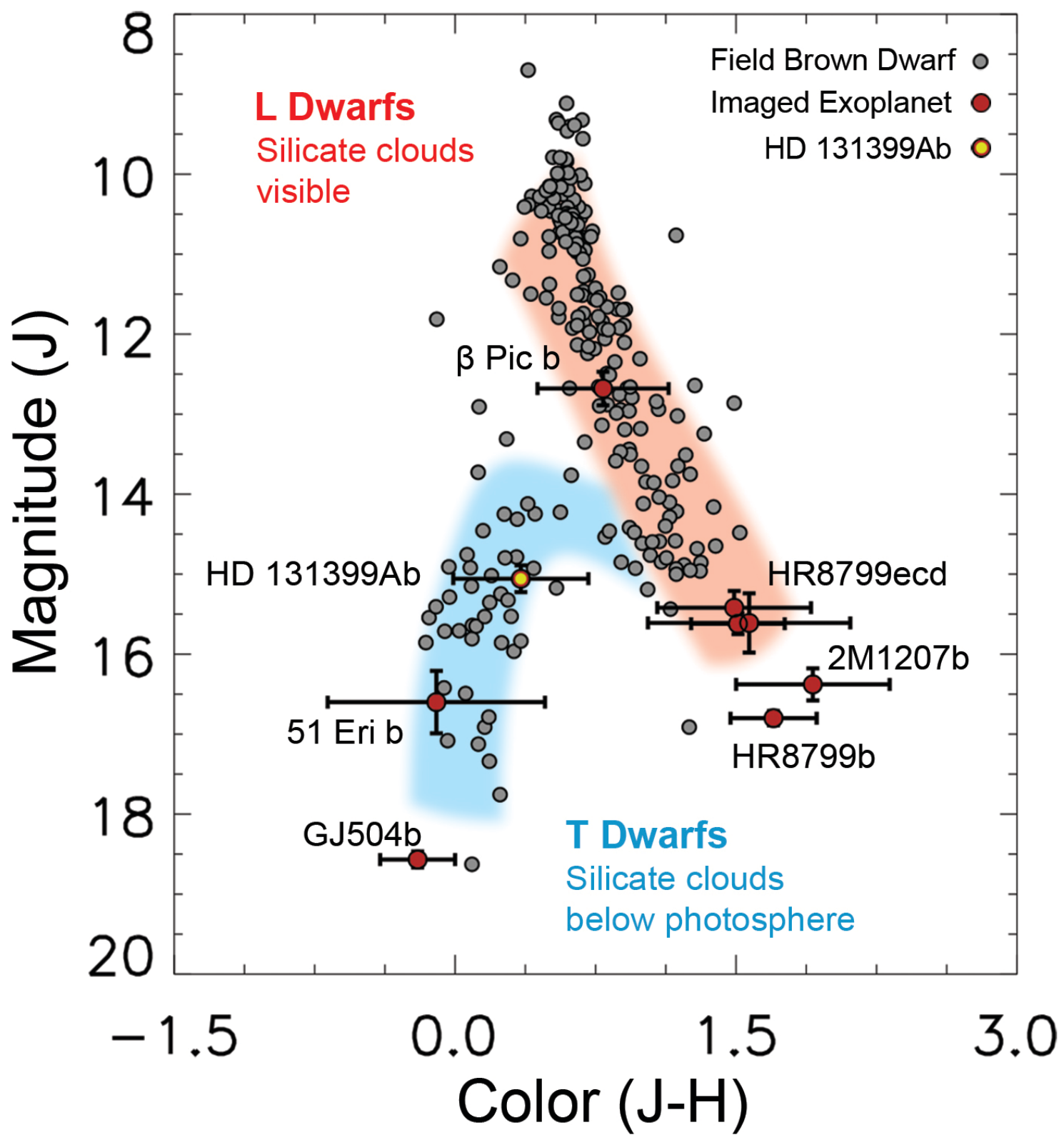

**Fig. 3.** ***J–H* color-magnitude diagram of brown dwarfs and directly imaged giant exoplanets.** HD 131399Ab falls among the methane dominated T dwarfs near the L/T transition.

The L and T dwarf data (with parallax calibrated absolute magnitudes) were obtained from (*41*), while the directly imaged exoplanet data are from (*16, 42-47*).

## Orbital Characterization of HD 131399

HD 131399Ab is the widest known exoplanet that orbits within a triple system (see Fig. 4, 6). Because the presence of a second and third star can greatly limit the phase space where planetary orbits are stable, observing a system in this configuration is thought to be unlikely (e.g. *7, 25*). In our ongoing survey, we have imaged 18 single A-type stars and 15 binary or triple star systems with separations similar to HD 131399A-BC. Although the sample size is small, it is surprising to us that the first planet detected in our survey is in a triple system.

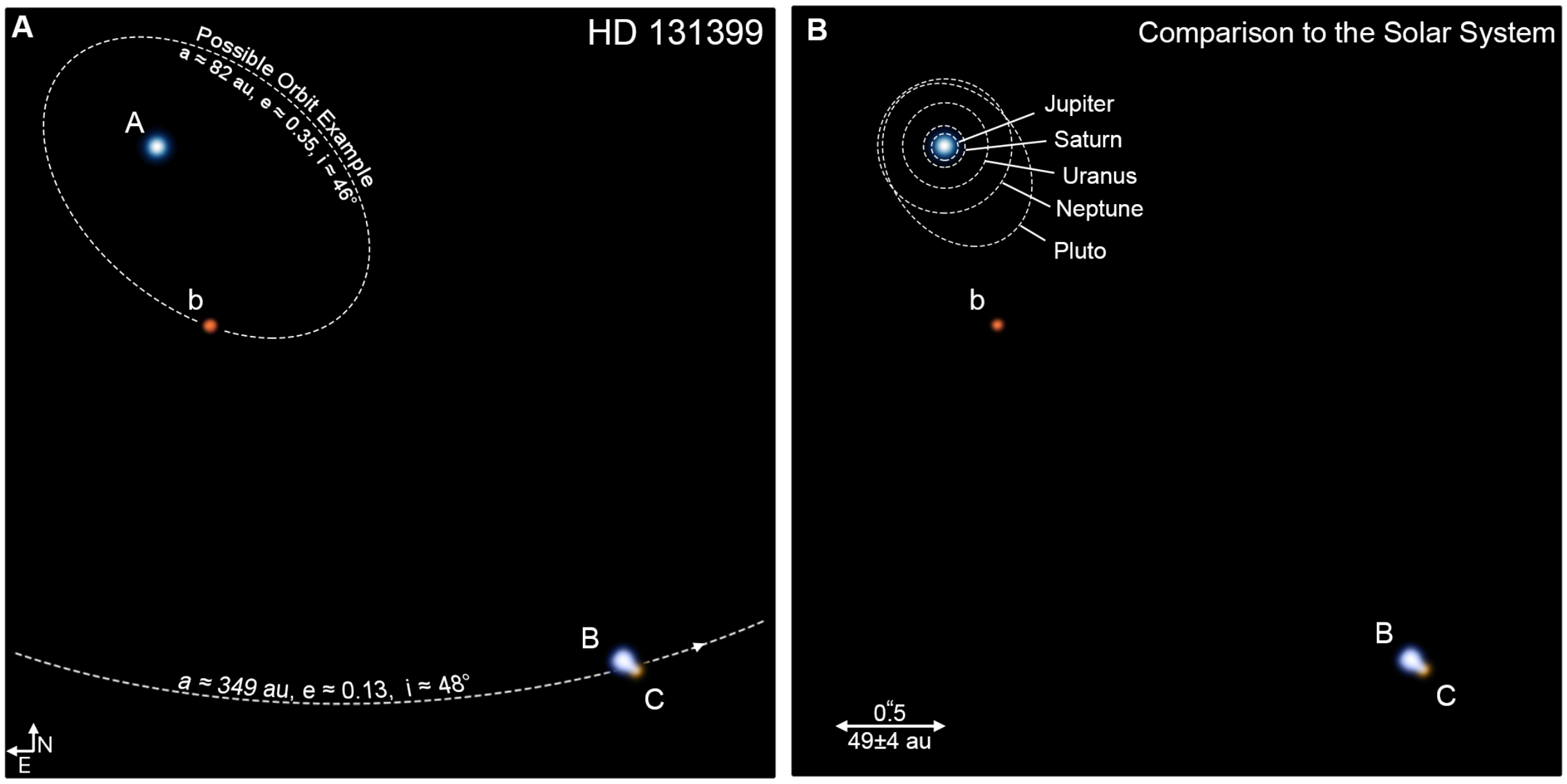

**Fig. 4: Schematic illustration of the components of the HD 131399 hierarchical triple star system and comparison to the Solar System.** Panel A: The dashed ellipses show our best-fit orbit of the BC pair and a preliminary orbit for the planet. The orbit shown for the planet has orbital elements that are consistent with the data, although the astrometric uncertainties permit a

significant range of orbits, with the parameter ranges given in the supplementary online text. Panel B: The image is reproduced with the orbits of the solar system planets overlaid. The underlying image is a composite image of the actual point-spread functions superposed on a dark sky background. The image is composed of SPHERE *J*–, *H*–, and *K*-band PSFs for components A and Ab (colored as blue, green, and red, respectively) and the monochromatic *K*-band PSF of components B and C. For clarity, the luminosity of the planet is enhanced by a factor of $10^5$, and since only *K*-band photometry exists for B and C their colors here are adjusted to be representative of typical G and K stars.

We use astrometric observations dating back to 1897 (Table S2; *26*) to fit the orbit to a grid of models for a binary system using the center of mass of the BC system. Our neglect of the BC orbit is motivated by the system's hierarchical nature and the fact that most previous data could not resolve the pair. Our best-fit model (Fig. S2, Table S3) consists of a semi-major axis of $a_\star = 349 \pm 28$ au, eccentricity of $e_\star = 0.13 \pm 0.05$, and inclination of $i_\star = 45°– 65°$ with respect to the plane of the sky, where the $\star$ subscript denotes the values for the BC orbit around A. Using only the newer, more reliable data permits a wider range of $a_\star = 270–390$ au, $e_\star = 0.1–0.3$, and $i_\star = 30°– 70°$. The available astrometry for the planet does not permit a robust orbital solution, though we perform a preliminary orbit-fit to obtain the plausible parameter ranges of $a_{\rm planet} = 82^{+23}_{-27}$ au, $e_{\rm planet} = 0.35 \pm 0.25$, and $i_{\rm planet} = 40^{+80}_{-20}{}°$, with no single solution being strongly preferred.

The orbital configuration of HD 131399 results in a more dynamically extreme configuration than for any known exoplanet within a binary or multiple system (Fig. 5, Table S4), with the ratio of semi-major axes $q = a_{\rm planet}/a_\star = 0.14–0.38$. Values of $q<0.23$ require higher eccentricities ($e_p>0.3$) to maintain the ≥82 au observational constraint on the planet's projected separation. The most dynamically similar planets to HD 131399Ab are the radial-velocity discovered γ Cep Ab (*27*), HD 41004Ab (*28*) and HD 142Ac (*29*) for which $q \sim 0.1$. Perhaps the most similar well-studied example is the transiting system Kepler-444, which hosts five sub-Earth sized planets within 0.1 au from the primary Kepler-444A (*30*). The latter stellar system is likewise a hierarchical triple, with a tight M-dwarf binary at 66 au from the planet hosting primary star. While similar to these other systems, HD 131399 stands out due to the proximity of the planet's orbit to that of the other stars in the system.

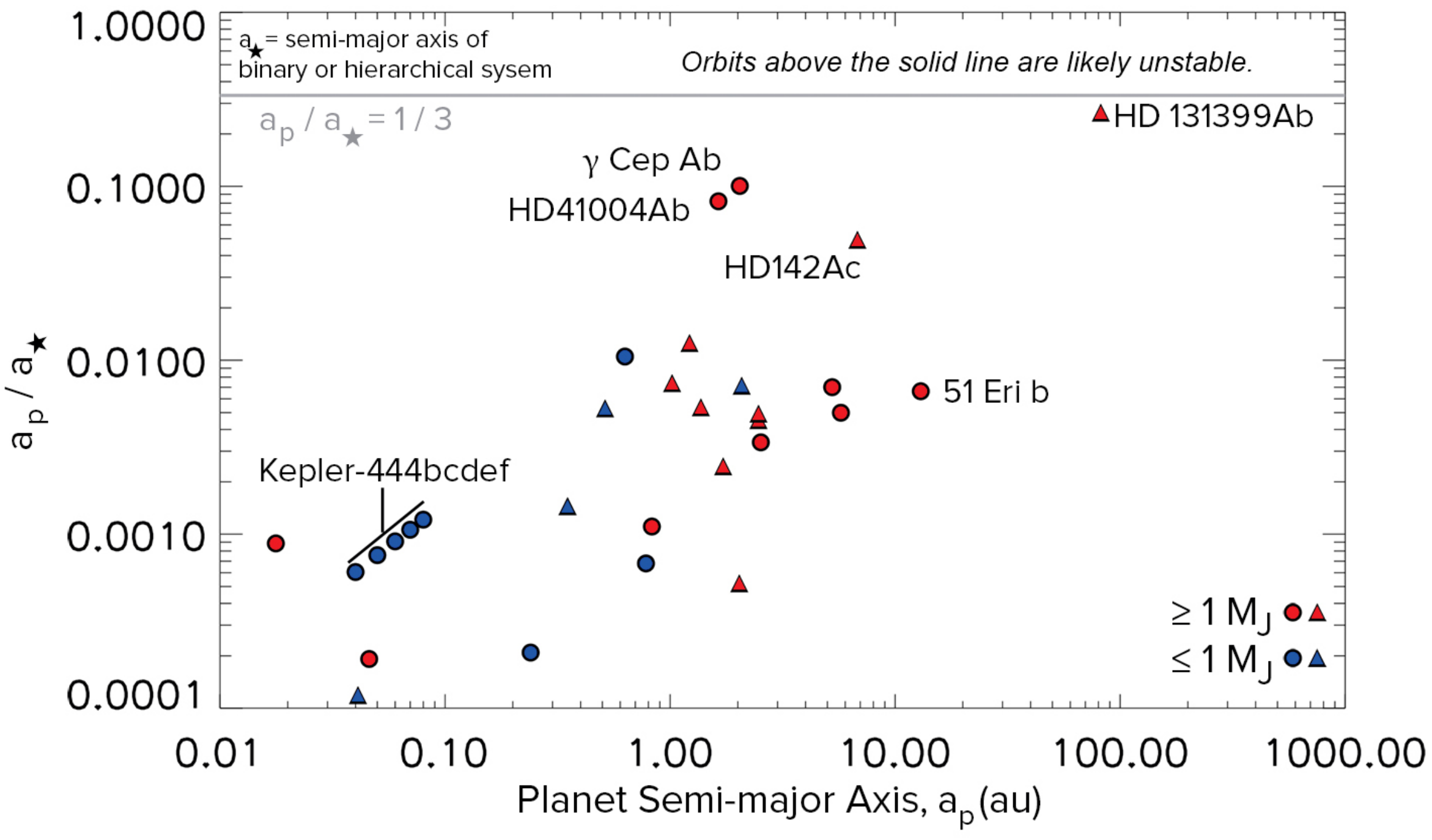

**Fig. 5: Ratio of semi-major axes of planets that orbit one star of a multiple system (satellite, or S-type planets) to the semi-major axes of their host systems.** The line at one-third times the binary separation represents the approximate critical radius of tidal truncation and orbital stability in the coplanar case *(25).* Although the critical radius varies somewhat for different parameters of stellar mass ratio, eccentricity, and inclination, HD 131399Ab is much closer to the critical radius than any other known exoplanet. For systems where either the planet or stars

lack precise orbital solutions, their projected separations are plotted instead – identified by triangle plot points instead of circles. This includes HD 131399Ab, although from the results of the preliminary orbit fit the semi-major axes of this system are indeed similar to the projected separations. See Table S4 for the list of included objects and their associated references.

We use a small suite (~300) of N-body simulations (*10*) to demonstrate that there exist stable orbital configurations for all four bodies that are consistent with the astrometric constraints. This holds even for the some of the more extreme configurations (i.e. smaller BC semi-major axis and higher eccentricity). The current astrometry also permits unstable orbits for the planet. Given the young age of the system, the planet might be on an unstable orbit, perhaps due to planet-planet or planet-star scattering, and could yet be ejected to become a free-floating planetary mass object. This is not the most likely scenario, as the timescale for the planet to suffer an ejection or collision is only a few Myr (*25*). In all cases, the orbit of b is non-Keplerian, as the planet's orbital parameters (*a,e,i*) undergo complex evolution due to the influence of the BC pair (Fig. S3).

**Formation of HD 131399Ab and the Origin of its Long-Period Orbit**

Given its location in a triple system, a broad set of formation pathways are possible for HD 131399Ab. It is unlikely that HD 131399Ab formed in isolation on its present long-period orbit around HD 131399A and is now on a stable orbit around HD 131399A, since planet formation is inhibited in the outer disk regions due to the strong perturbations from the binary (*e.g. 31, 32*). We speculate that the planet may have arrived at its present orbit through one of three possible scenarios: Scenario A) the planet formed on a short orbit around star A, and subsequently underwent a planet-planet scattering event that ejected it to its current long-period orbit (*33*). This scenario requires the presence of a massive planet on a shorter period orbit. Such a planet could have evaded detection if it were beneath our sensitivity limits (see supplementary online text for details). As a consequence we would also expect the Ab orbit to be rather eccentric. Scenario B) HD 131399Ab formed as a circumbinary planet around components B and C and underwent a scattering event via interactions with another planet or with the binary itself (*6*). This scenario would also be most consistent with an eccentric Ab orbit. Scenario C) the planet formed around either component before the A-BC system arrived in its present configuration. The stellar orbits could have evolved subsequently due to interactions with the natal disks or secular effects (*34*). This scenario does not require the presence of a second close-in massive planet, though the resulting outer planetary orbit may be indistinguishable. Thus it is possible that the planet is no longer orbiting the star around which it formed. These scenarios are also consistent with Ab obtaining an orbit around all three components, although the short lifetime of such an orbit makes this configuration unlikely.

**Acknowledgments:** This work is based on observations performed with VLT/SPHERE under program IDs 095.C-0389A (PI: D. Apai) and 296.C-5036A (PI: K. Wagner). KRW is supported by the National Science Foundation Graduate Research Fellowship Program under grant No. 2015209499. The results reported herein benefited from collaborations and/or information exchange within NASA's Nexus for Exoplanet System Science (NExSS) research coordination network sponsored by NASA's Science Mission Directorate. This research has benefitted from the SpeX Prism Spectral Libraries, and the Washington Double Star Catalog maintained by the U.S. Naval Observatory at http://www.usno.navy.mil/USNO/astrometry/optical-IR-prod/wds/WDS. All atmospheric models used in this study can be found on online at http://svo2.cab.inta-csic.es/theory/newov/ and all of the raw data products for HD 131399Ab and associated calibrations may be obtained from the ESO archive at http://archive.eso.org/cms/eso-data.html.

SPHERE is an instrument designed and built by a consortium consisting of IPAG (Grenoble, France), MPIA (Heidelberg, Germany), LAM (Marseille, France), LESIA (Paris, France), Laboratoire Lagrange (Nice, France), INAF - Osservatorio di Padova (Italy), Observatoire de Genève (Switzerland), ETH Zurich (Switzerland), NOVA (Netherlands), ONERA (France) and ASTRON (Netherlands) in collaboration with ESO. SPHERE was funded by ESO, with additional contributions from CNRS (France), MPIA (Germany), INAF (Italy), FINES (Switzerland) and NOVA (Netherlands). SPHERE also received funding from the European Commission Sixth and Seventh Framework Programmes as part of the Optical Infrared Coordination Network for Astronomy (OPTICON) under grant number RII3-Ct-2004-001566 for FP6 (2004-2008), grant number 226604 for FP7 (2009-2012) and grant number 312430 for FP7 (2013-2016).

**Supplementary Materials:**

Materials and Methods

Supplementary Text

Figures S1-S4

Tables S1-S4

References (*48*-*83*)

# Supplementary Materials for

## Direct Imaging Discovery of a Jovian Exoplanet Within a Triple Star System

Kevin Wagner, Dániel Apai, Markus Kasper, Kaitlin Kratter, Melissa McClure, Massimo Robberto, Jean-Luc Beuzit

Correspondence to: kwagner@as.arizona.edu

**This PDF file includes:**

Materials and Methods
Supplementary Text
Figs. S1 to S4
Tables S1 to S4

**Materials and Methods**

VLT/SPHERE Description:

Our observations were carried out with SPHERE, the Spectro-Polarimetric High-contrast Exoplanet Research instrument (*9*), which provides high imaging contrast by combining extreme adaptive optics (*48*), coronagraphy, and accurate calibration of non-common path instrumental aberrations. Three scientific cameras record and analyze the high-contrast images: the differential near-infrared (*Y-K* bands) imaging camera and spectrograph (IRDIS, InfraRed Dual Imaging Spectrograph; *49, 50*); the near-infrared (*Y*-H bands) low spectral resolution Integral Field Spectrograph (IFS, *21*); and the visible (500-900 nm) imaging differential polarimeter (ZIMPOL, Zurich Imaging Polarimeter, *51*). With these, SPHERE provides imaging contrasts of one part in a million at angular separation of a few tenths of an arcsecond.

Observing Log and Data Reduction:

The basic details of the observing runs are given in Table S1. We reduced the IRDIS data using the standard European Southern Observatory (ESO) data reduction pipeline for SPHERE *(52),* which includes dark subtraction, flat field division, and bad pixel correction. For the IFS data we used the pipeline described in *(53),* which performs the same operations described for IRDIS along with an additional wavelength calibration step and spectral cross-talk correction. The data were corrected for field distortion *(54)* and were spatially filtered by a high-pass filter to remove the stellar halo and any remaining low-frequency flat-field variations. The filtered images are useful for the initial detection of point sources, though the data were reprocessed without this step to preserve photometric accuracy.

HD 131399Ab was detected as a distinct point source following PSF subtraction using a principal component analysis (PCA) based Karhunen-Loève Image Processing (KLIP) procedure *(55)* operating in angular differential imaging mode (ADI), in which the intrinsic field rotation of the Altitude-Azimuth telescope mount is used to distinguish the rotating flux of the planet from the pupil-stabilized PSF of the telescope aperture. Following ADI processing, we further reduced the speckle noise in the IFS data by performing a second KLIP step in spectral differential imaging (SDI) mode, where the radial scaling of the PSF with wavelength is utilized for its reconstruction and subtraction within each temporal data cube.

We performed numerous trials with various input parameters for the KLIP algorithm and find an optimal throughput of the planet flux by retaining the first four principal components in the PSF reconstruction, where the analysis is optimized individually for annular segments of 60° arc-length and 14 pixel width. In the ADI (SDI) step we excluded frames whose natural field rotation (wavelength scaling) is within 1.5 times the FWHM to avoid self-subtraction of the planet. With the combined ADI+SDI method a higher signal to noise ratio may be achieved than with either method alone. The detection of HD 131399Ab is apparent and grows in significance with each step. Since the SDI method introduces strong correlation between the individual spectral channels, the spectrum and photometry of HD 131399Ab were extracted from the ADI-only processed images, although the systematic errors of neighboring channels are still correlated due to the similar residual speckle patterns of neighboring wavelengths. For a detailed discussion of these effects, see (*56*).

We measured the throughput of astrophysical point sources through the KLIP reduction by injecting synthetic planets of known contrast and position into the raw data and found the throughput to be near unity (within 5% on average in all wavelengths) for point sources at the separation of HD 131399Ab for our chosen reduction parameters. As a result, no additional throughput correction has been performed, though this additional uncertainty is accounted for in the values reported herein. The throughput decreases for reconstructions including more than the first 4–5 principal components, which supports our choice of a non-aggressive ADI including only the first four principal components. We performed a second analysis of the number of principal components that maximizes the signal to noise ratio (*SNR*), and find a maximum *K1*-band *SNR*=23.5 for HD 131399Ab with 15 principal components used in the reconstruction, as opposed to *SNR*=19.9 with the utilization of only the first four components. However, this more aggressive step reduces the throughput to 70-80% with a significant (10%) variance with position angle. The photometry and spectra were extracted from the non-aggressively reduced data to avoid introducing this larger throughput uncertainty.

Astrometric Calibration:

The ESO standard calibration plan for SPHERE includes observations of astrometric reference fields every one to two weeks to ensure astrometric accuracy. The closest calibrations to the dates of our own runs observed an astrometric reference globular cluster, NGC6380, on 2015-06-02, 2015-07-09, 2016-03-06, and 2016-05-04. At a distance of 10.5 kpc, the low relative motion of the stars in NGC6380 allow for precise determination of the field orientation and plate scale that are necessary for precision astrometry across multiple epochs. To calibrate the SPHERE images, we downloaded archival *Hubble Space Telescope* WFC3/UVIS data (PI: Noyola, 2010-03-09) which is properly calibrated to within 0.1 degree of the International Celestial Reference System North-pole (*57*).

We measured the position angle and separation for pairs of the nine brightest stars in the ~6 arcsec vicinity of HD 159073 (which is itself saturated in the HST images, and therefore excluded from our measurements) with respect to the pre-calibrated "North" of the image plane. We performed an identical analysis on the SPHERE and HST images, and find a median difference in field rotation of $-1.57 \pm 0.2$ degrees for 2015-06-02, $-1.55 \pm 0.2$ degrees for 2015-07-09, $-1.40 \pm 0.2$ degrees for 2016-03-06, and $-1.47 \pm 0.2$ degrees for 2016-05-04 where the uncertainty is in equal parts due to the precision of the HST alignment and from the standard deviation (~0.1 degree) of the measurements of the individual stellar pairs. In a similar fashion, we obtain plate scales for IRDIS images of 12.23±0.03 mas/pixel for 2015-06-02 and 12.24±0.03 mas/pixel for subsequent epochs. We tested our method by introducing randomly assigned rotations and found that we are able to recover the true orientation to within the desired 0.1 degree accuracy and to within 0.01 degrees in most cases. After including all nine stars, we repeated the analysis with only the brightest five stars to reduce errors due to the larger uncertainty in determining the centroid of the fainter stars, and found that the two methods are consistent within ~0.01 degree.

An additional star-centering calibration step was executed during each observing block in order to accurately determine the position of the primary star behind the coronagraph. During this calibration a waffle pattern is applied to the deformable mirror, casting satellite spots of starlight at specific positions in a cross pattern surrounding the star. The positions of these spots are used to determine the geometric location of the star behind the coronagraph before and after the observations, and an interpolation is done to center the frames in between. Errors in interpolation (e.g. non-linear drift) would not have affected our final results since the difference in star position before and after the observing blocks typically differs only by half of a pixel or less. It is also unlikely that the tracking drifted and then returned to its original position during our observations since SPHERE is running a closed loop on a dedicated tip tilt sensor close to the coronagraph's focal plane, which typically leaves residual drifts < 1-2 mas rms. The positions of the brighter components B and C were extracted from the non-saturated flux calibration images, in which the primary star is slewed off of the coronagraph and hence its position can be measured in the traditional manner of determining its photometric centroid.

We extracted the position of the planet at each epoch via injection of a negative Gaussian PSF over a grid of 0.01-pixel spacing, from which we determined its position as the location with the minimum square of residual intensity in an aperture of one full width half maximum diameter around the planet. This method is typically accurate to determine the center of a well-sampled PSF to within ~0.1 pixels (*58*), which enables precision astrometry for the planet's confirmation and orbital analysis.

Age Analysis:

Since planets become less luminous as they age, the mass estimates of directly imaged planets (typically estimated from their luminosity) rely heavily on the ages of their host systems. There are unfortunately few age indicators available for A-type stars, making age estimates difficult for A-stars that aren't members of any known association. As a member of the Upper-Centaurus-Lupus (UCL) association, the age of HD 131399 is uniquely well constrained compared to most directly imaged exoplanet host stars, which do not belong to similarly well-dated associations. Based on the star's distance and proper motion, its probability of UCL membership has been established with 94% confidence via the *Hipparcos* mission (*11*). This result has been independently confirmed via multi-dimensional Bayesian analysis of the star's kinematics (*59*), resulting in a membership probability higher than 91% – among the highest probability reported therein for any of the Hipparcos stars typically considered as members of UCL. Since no other data currently exist that could constrain the age of the system (e.g. stellar rotation period or multi-band photometry of the lower-mass components B and C), the membership in UCL is presently the best constraint on the age of the system. Additional estimates on the system's age are beyond the scope of our present study.

Orbital Characterization of HD 131399A-BC:

Although HD 131399ABC is a triple system, its hierarchical nature allows us to fit the orbit of the BC center of mass about A from over one hundred years of astrometric measurements (Table S2). Using the method described in (*60*) to calculate the position angle (PA) and separation of a pair of binary stars as a function of time and orbital parameters, we minimized the residuals in a reduced chi-squared sense over a grid of models to find the best-fit parameters (Table S3). We fixed the total mass of the system to be the sum of the masses of the three stellar components in order to eliminate period as an independent variable in favor of semi-major axis. In measurements where B and C are resolved, we use their barycenter as the relevant position measurement. Our routine performs seven iterations of parameter searches over a uniformly spaced grid of 46,656 combinations of the six orbital parameters, where at each iteration the grid range and spacing are reduced by 50%. To assign ranges to the fit parameters we repeated the analysis for thousands of parameter retrievals, each time modulating the data by their uncertainty multiplied by random Gaussian noise, to build up a range of best-fit parameters corresponding to the level of uncertainty in the data.

We first fit only the data after 1990 due to the lack of documented uncertainties in the earlier data, although this approach presents only seven data points and very little motion to constrain the orbit of the system. The profile of these parameter retrieval distributions is non-Gaussian, and hence we quote the range of parameters retrieved in Table S3 without quoting a best-fit orbit. To make use of the older available data, we repeated the analysis with estimated uncertainties assigned to the older data by extrapolating the trend in uncertainties back in time, arriving at $\Delta\rho = 0.9$ arcsec and $\Delta\theta = 22°$ for the 1902 data, which we find to be both reasonable and conservative estimates for the technological and atmospheric limitations of the time. We chose to exclude the first measurement taken in 1897 as an outlier due to its large position angle difference with the other measurements of the same era. We fit the resulting arrays of best-fit parameters to normal distributions and from these we assign the true value of the orbital parameter to be the mean and its uncertainty to be the half-width of the 95% confidence interval. Table S3 shows the values of the best-fit parameters, and Fig. S2 shows a sample of potential orbits spanning the approximate range of the fit uncertainties. Though the data do not allow for precise determination of the system's orbital properties, the rough constraints placed with this initial body of data importantly affirm the conclusion that the stars in the HD 131399 system are close enough in semi-major axis to the planet to be of dynamical significance.

N-body Simulations and Orbital Stability Analysis

To explore the stability and orbital evolution of the system, we use the N-body code Swifter (*61*) to integrate the orbits of all four bodies (three stars, one planet). We use the Radau integrator, which is well suited to hierarchical, multi-body systems. Because the current astrometric constraints, especially for the Ab component, are limited, we only perform a sparse sampling of the parameter space. We ran roughly 300 models varying the binary semi-major axis (300-310 au) and eccentricity (0-0.4), the planet's eccentricity (0-0.75) and mass (4, 12 $M_{Jup}$), and the relative inclination (0-180 degrees). We fix the apocenter of Ab at the projected separation of 82AU. Models are integrated out to 100 Myr. We find that stable orbits exist for Ab even for the shorter and more eccentric A-BC orbits consistent with our fits ($a_{AB}$ ~ 300AU, $e_{AB}$ ~ 0.2). The A-BC orbit is stable independent of the parameters, as the planet's mass constitutes a small perturbation. In general, higher planet eccentricity and/or higher inclination orbits are more stable, as these reduce the typical closest approach distance of the planet with the BC pair. The mass of the planet has little effect on stability.

## Supplementary Text

Additional Dynamical Effects from HD 131399C and Other Stars

To our knowledge, this is the first detection of HD 131399C. Though we lack a sufficient baseline to constrain its orbital properties, we confirm proper motion with the system and from its close proximity (~7.5 au) to HD 131399B we infer the two to be a tight binary. Due to the system's hierarchical nature the existence of HD 131399C is comparable to the scenario of a star with the combined mass of stars B and C at the location of their barycenter.

The Washington Double Star Catalogue lists an additional and different component 'C' of the system, an A3III pre-main sequence star at 33 arcseconds from HD 131399A. However, its change in separation with HD 131399A of 3.2 arcseconds over the last century is an order of magnitude higher than expected for a bound orbit even in the most favorable case, while consistent with the expectation for an unbound object whose apparent proximity is due to a chance alignment. As a result we expect no additional dynamical consequences from this star, unless it has undergone a recent close encounter with the HD 131399 system.

Sensitivity and Limits on the Presence of Other Planets

We have assessed the sensitivity of our observations by injecting fake planets (PSFs) into the individual *K1* images, where the contrast with the star is most favorable. The injected planets are of known contrast and location, allowing for characterization of the achieved contrast of the observations as a function of separation from the star. Following angular differential imaging via the KLIP PCA-based method (*55*), we find that our observations are sensitive to contrasts of $10^{-4}$ at 0.25 arcsec, $10^{-5}$ at 0.45 arcsec, and $10^{-5.5}$ at 0.65 arcsec (Fig. S4). None of the injected planets with a contrast of $10^{-6}$ were recovered. Through comparison to hot-start models we convert these sensitivities (as absolute magnitudes) to upper mass limits of ~10 $M_{Jup}$ at 0.25 arcsec (25 au), ~5 $M_{Jup}$ at 0.45 arcsec (44 au), and ~3 $M_{Jup}$ outside of 0.65 arcsec (64 au), where we have conservatively rounded each mass estimate upward instead of to the nearest digit. While there are a number of other tentative point source detections, none are confirmed across multiple epochs, and are likely residual speckles. We do not detect an inner planet, whose existence would help to explain how HD 131399Ab arrived at its present long-period orbit, nor do we detect an outer planet within our 11” × 12.5” field of view.

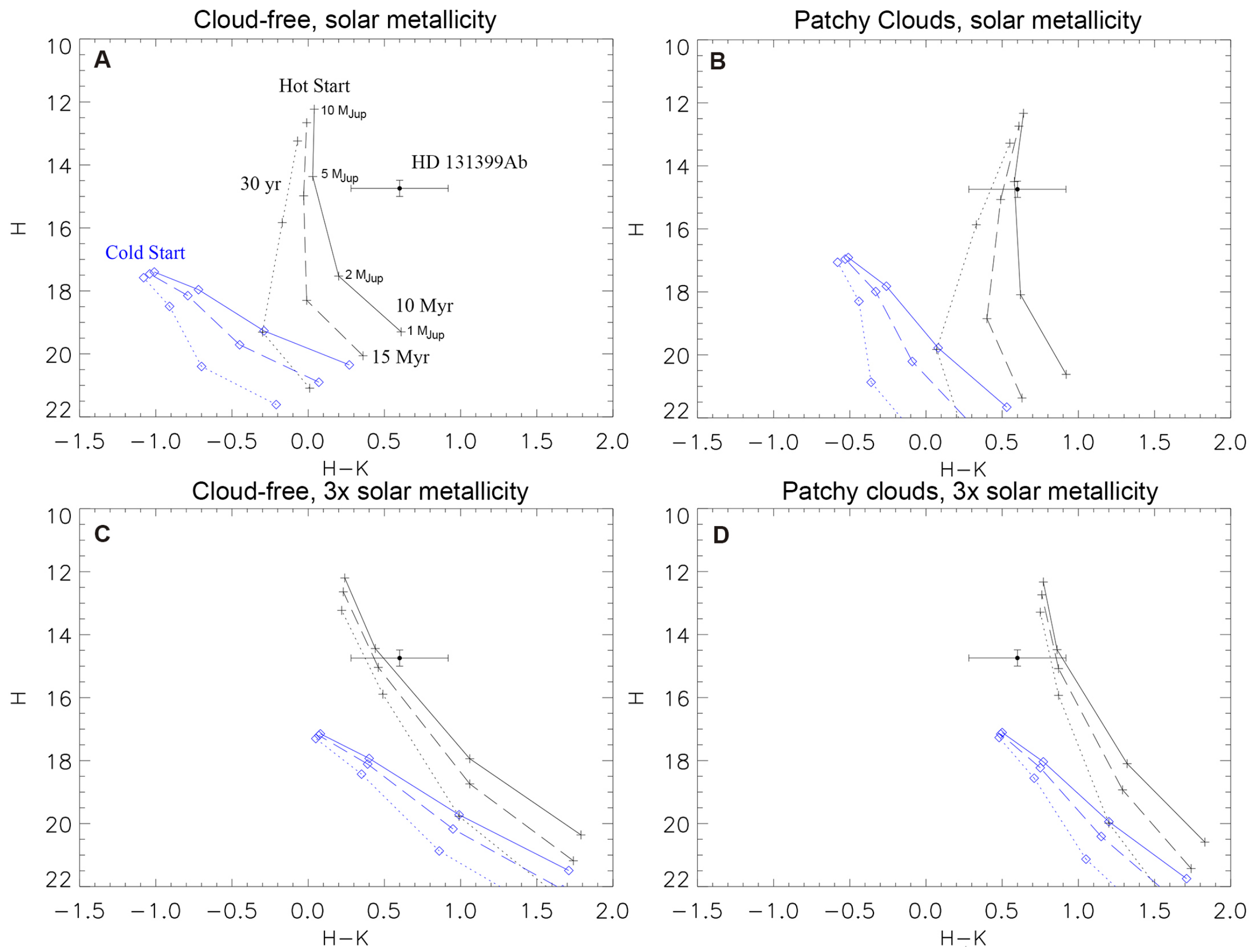

**Fig. S1. Color-magnitude comparison of *H*–, and *K*-band aperture photometry.** The data are compared to the hot start (high entropy initial conditions) and cold start (low entropy initial conditions) models from (*19*) with different atmospheric properties in panels A, B, C, and D. The plotted symbols represent masses of 10, 5, 2, and 1 $M_{Jup}$, respectively from brightest to faintest.

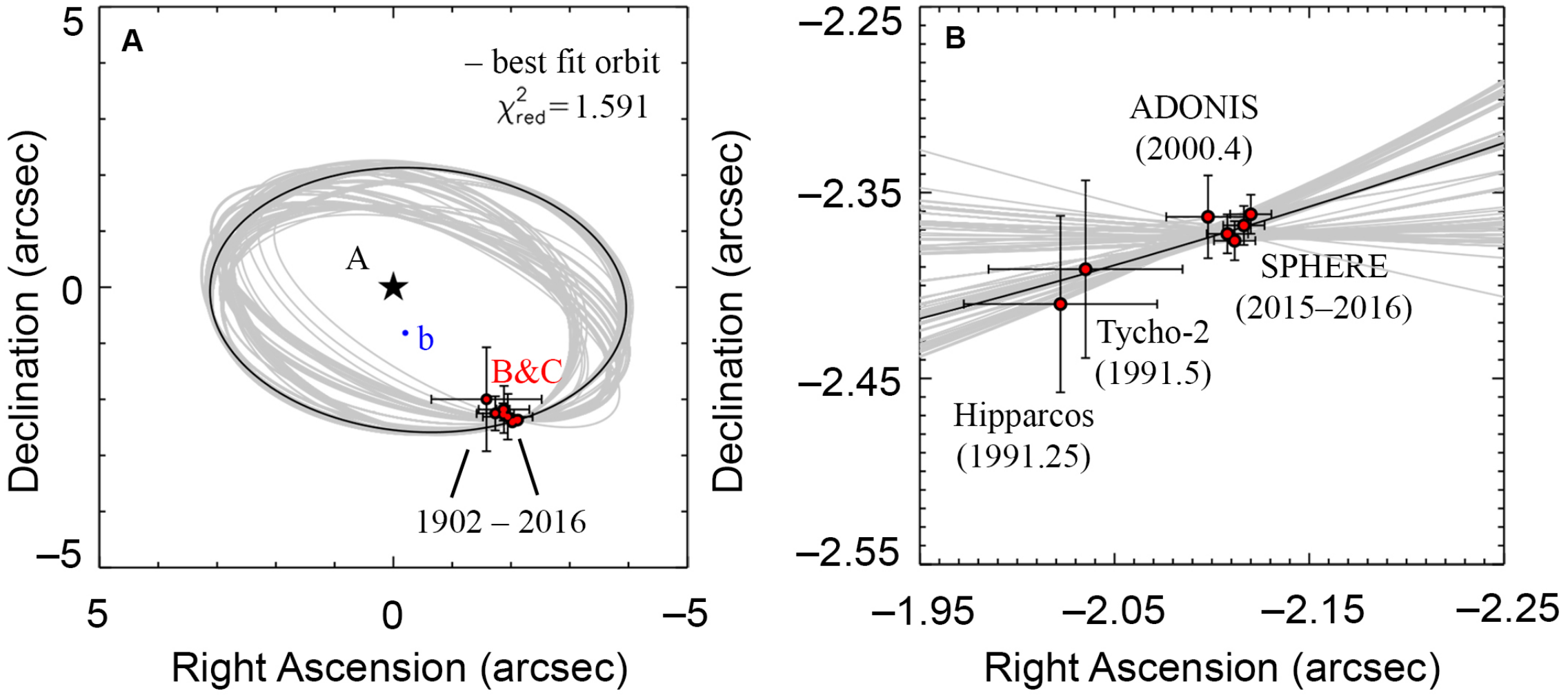

**Fig. S2. Best-fit orbital models of HD 131399A-BC.**

Model orbits are shown for the orbital parameters given in Table S3, with the best-fit model shown in black and a sample of 100 randomly selected orbits within the model uncertainties shown in gray. The orbits shown here were fit to the full 1902–2016 data set and are shown as they would appear in the plane of the sky. Their departure from circularity is due to a range of eccentricities and inclinations. The orbits are clustered around e~0.1 and a~350 au, suggesting the pair is indeed close enough to be of significant dynamical influence to the planet at ~82 au. Panel A shows the full orbital range, while panel B shows a close-up of the more recent data from 1991–2016. The positions are given as offsets with the position of HD 131399A at each epoch as the zero-point.

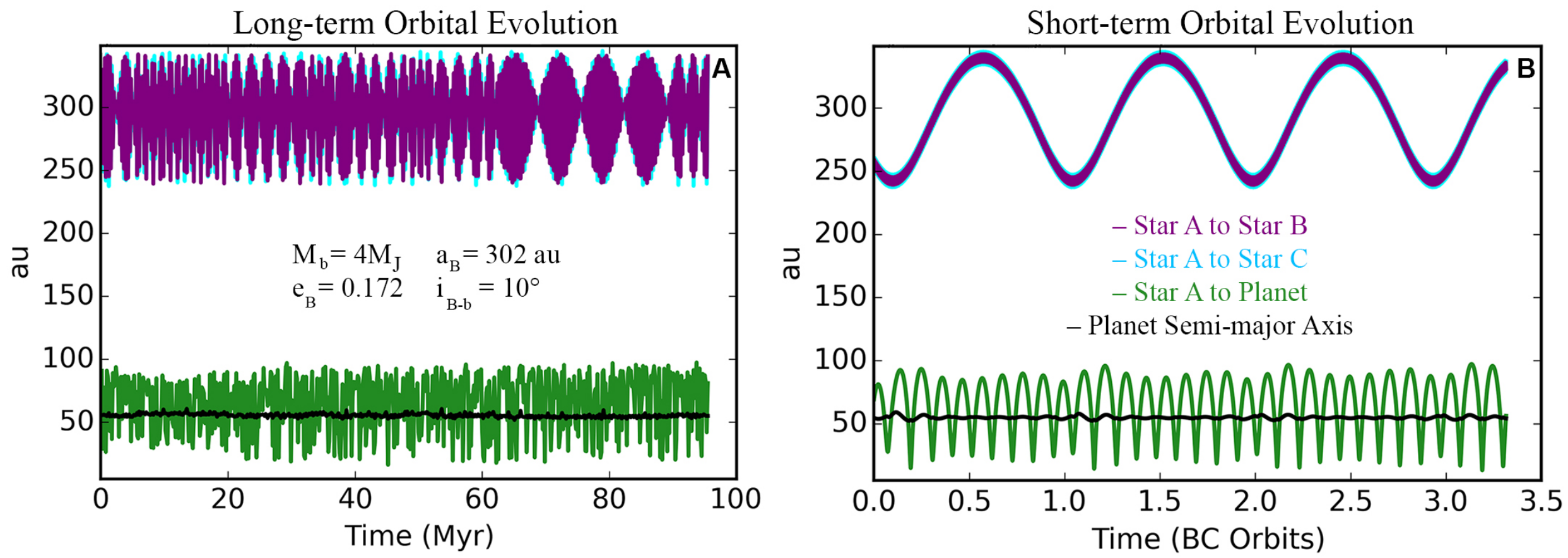

**Fig. S3. An example of short-and long-term orbital evolution of the HD 131399 components.** The orbits for all four components are stable for 100 Myr and the orbit of component Ab undergoes both short and long timescale variations due to perturbations from BC, though it is not necessary that the planet is on a stable orbit given the system's youth. Note that the osculating orbital elements for all bodies can show large amplitude, short timescale variations because the orbits are non-Keplerian. Panel A shows the orbital evolution for one of the possible stable orbital configurations over 100 Myr, while panel B shows the short-term evolution over the period of the A-BC orbits.

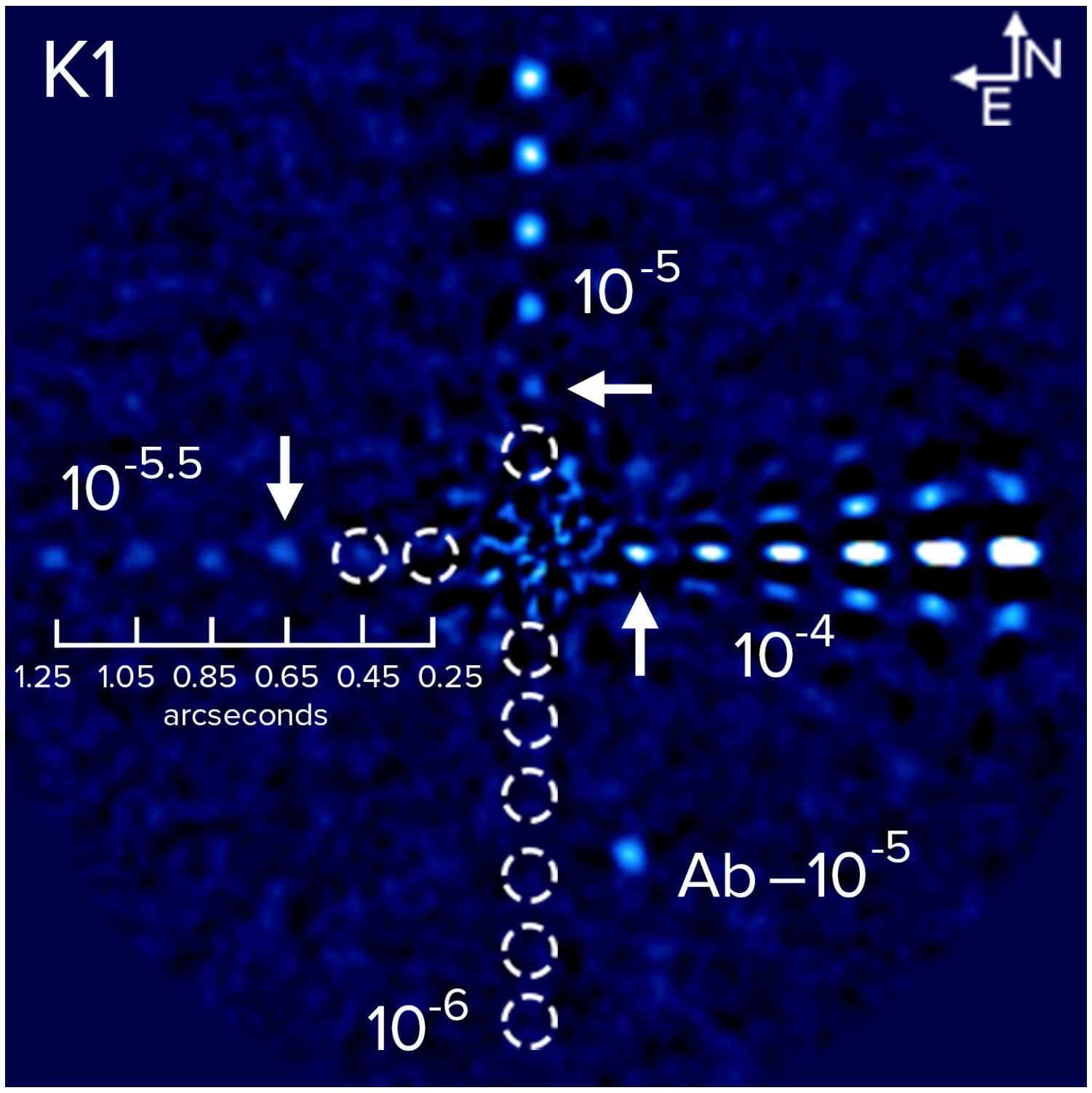

**Fig. S4. Contrast sensitivity assessment via synthetic planet injections.**

Six synthetic planets were injected along each of the cardinal axes, with uniform contrasts along each axis to the star (labeled in the image). The arrows indicate the innermost planets of each contrast that would have been identified as planet candidates in our observations, while the locations of unrecovered planets are identified in dashed circles. HD 131399Ab itself appears at a contrast of $10^{-5}$. None of the planets injected along the South axis with a contrast of $10^{-6}$ were recovered. The bright spots on either side of the synthetic planets with $10^{-4}$ contrast are artifacts of the ADI data reduction process.

| Obs. Date | Program ID | Instrument Mode: Filters | Int. Time (s) (IRDIS, IFS) | Seeing (arcsec) | Airmass | Field Rot. ($^o$) |
|---|---|---|---|---|---|---|
| 2015 June 12 | 095.C-0389A | IFS: YJH IRDIS: K1&K2 | 1536, 1600 | 0.9 | 1.02 | 36.9 |
| 2016 March 06 | 296.C-5036A | IFS: YJH IRDIS: K1&K2 | 1792, 1792 | 1.2 | 1.03 | 38.0 |
| 2016 March 17 | 296.C-5036A | IFS: YJH IRDIS: K1&K2 | 1792, 1792 | 1.1 | 1.08 | 37.2 |
| 2016 April 02 | 296.C-5036A | IFS: YJH IRDIS: K1&K2 | 224, 224 | 0.8 | 1.02 | 4.3 |
| 2016 May 07 | 296.C-5036A | IFS: YJH IRDIS: K1&K2 | 1792, 1792 | 0.9 | 1.01 | 40.4 |

**Table S1. Observing log of HD 131399 with VLT/SPHERE.**

The 2016 April 2 observations were aborted due to changing weather conditions and the low cadence yielded inadequate field rotation for the detection of the planet, though the data are of sufficient cadence for astrometric measurements of the stellar components. The rest of the data were combined to produce the spectrum and photometry of HD 131399Ab presented herein.

| **Date (J2000)** | **Separation (arcsec)** | **Position Angle (°)** | **Reference** |
|---|---|---|---|
| 1902.40 | 2.6±0.9 | 218±22 | (*62*) |
| 1927.01 | 2.9±0.4 | 221±9 | (*63*) |
| 1928.56 | 3.0±0.4 | 220±9 | (*64*) |
| 1939.79 | 2.8±0.3 | 217±7 | (*65*) |
| 1965.5 | 2.9±0.1 | 219.5±2.5 | (66) |
| 1966. | 2.9±0.1 | 220.6±2.5 | (67) |
| 1991.25 | 3.146±0.042 | 220.0±1.0 | (*14*) |
| 1991.5 | 3.140±0.042 | 220.4±1.0 | (*68*) |
| 2000.43 | 3.160±0.025 | 221.6±0.3 | (*38*) |
| 2015.45 | 3.174±0.01 | 221.9±0.3 | This work |
| 2016.18 | 3.173±0.01 | 221.6±0.3 | This work |
| 2016.21 | 3.176±0.01 | 221.8±0.3 | This work |
| 2016.26 | 3.179±0.01 | 221.6±0.3 | This work |
| 2016.38 | 3.175±0.01 | 221.9±0.3 | This work |

**Table S2. Astrometry of HD 131399A-BC.**
For measurements where B and C are unresolved, their photocenter is used and the quoted uncertainties are large enough in each case to cover the difference between photocenter and barycenter of the tight binary. For our SPHERE observations, in which the two components are resolved, we report their computed barycenter and propagated uncertainties reflecting this calculation. For the older measurements, we've assigned conservative estimates for the uncertainties as described in the text.

| **Parameter** | **A-BC orbit (fit to 1991-2016)** | **A-BC orbit (fit to 1902-2016)** | **Preliminary Ab orbit** |
|---|---|---|---|
| Period | 2800–3600 years | 3556±36 years | 400–700 years |
| Time of periastron passage (AD years) | 200–500 AD | 502±33 AD | 1600–1950 AD |
| Eccentricity | 0.1–0.3 | 0.13± 0.05 | 0.35±0.25 |
| Semi-major axis | 3.0–3.7 arcsec (273–389 au) | 3.56±0.03 arcsec (349±28 au) | 0.6–1.0 arcsec (55–105 au) |
| Inclination | 30° – 70° | 45° – 65° | $40^{+80}_{-20}{}^{\circ}$ |
| Longitude of periastron | | 145.3°±15°, 310°±10° | |
| Longitude of ascending node | | 265°±20°, 75°±10° | |

**Table S3. Orbital Models for HD 131399A-BC and HD 131399A-Ab.**

For the best-fit parameters from 1902-2016 the value and uncertainty are given as the center and 95% confidence interval of the normal distribution of parameter retrievals, except for the parameters of longitude of periastron and longitude of ascending node, which follow double peaked distributions that are approximately 180° apart.

| Planet | $a_p$ (au) | $a_\star$ (au) | $a_p/a_\star$ | Orbital Solution? | Reference |
|---|---|---|---|---|---|
| HD 131399Ab | 82 | 309 | 0.265 | No | This work |
| γ Cep Ab | 2.04 | 20 | 0.102 | Yes | (*69*) |
| HD 41004Ab | 1.64 | 22 | 0.0745 | Yes | (*28*) |
| HD 142Ac | 6.8 | 106.1 | 0.0641 | No | (*29*) |
| HD 177830Ab | 1.22 | 97.1 | 0.0126 | No | (*70*) |
| Kepler-64b | 0.63 | 60 | 0.0105 | Yes | (*71*) |
| HD 142 Ab | 1.02 | 106.1 | 0.00961 | No | (*72*) |
| HD 114729Ab | 2.08 | 291 | 0.00715 | No | (*73*) |
| Ups And e | 5.25 | 750 | 0.007 | Yes | (*74*) |
| 51 Eri b | 13 | 1960 | 0.00663 | Yes | (*16*) |
| HD 65216Ab | 1.37 | 255.2 | 0.00537 | No | (*75*) |
| HD 177830Ac | 0.514 | 97.1 | 0.00529 | No | (*76*) |
| 55 Cnc d | 5.74 | 1150 | 0.00499 | No | (*77*) |
| HD 196050Ab | 2.47 | 501 | 0.00493 | No | (*72*) |
| Ups And d | 2.53 | 750 | 0.00337 | Yes | (*74*) |
| 16 Cyg Bb | 1.72 | 700 | 0.00246 | No | (*78*) |
| HD 16141Ab | 0.35 | 241.5 | 0.00145 | No | (*72*) |
| Kepler-444Af | 0.08 | 66 | 0.00121 | Yes | (*30*) |
| Ups And c | 0.83 | 750 | 0.00111 | Yes | (*74*) |
| Kepler 444Ae | 0.07 | 66 | 0.00106 | Yes | (*30*) |
| Kepler-444Ad | 0.06 | 66 | 0.000909 | Yes | (*30*) |
| Kepler-44Ac | 0.05 | 66 | 0.000758 | Yes | (*30*) |
| 55 Cnc f | 0.781 | 1150 | 0.000679 | No | (*77*) |
| Kepler-444Ab | 0.04 | 66 | 0.000606 | Yes | (*30*) |
| HD 213240Ab | 2.03 | 3888 | 0.000522 | No | (*79*) |
| 55 Cnc c | 0.24 | 1150 | 0.000209 | No | (*77*) |
| Tau Boo b | 0.046 | 240 | 0.000192 | Yes | (*80*) |
| HD 46375Ab | 0.041 | 345 | 0.000119 | No | (*81*) |
| 55 Cnc b | 0.11 | 1150 | 0.0000957 | No | (*77*) |
| Ups And b | 0.06 | 750 | 0.00008 | Yes | (*74*) |
| 55 Cnc e | 0.0156 | 1150 | 0.0000137 | No | (*77*) |

**Table S4. Separations of S-type planets ($a_p$) and stars ($a_\star$) in multi-star systems.** Where precise orbital solutions are available for both components, the semi-major axes are listed as the separation value. For systems lacking an orbital model (usually due to the long period of one component), the projected separations are given instead.